\documentclass{sigchi}

\usepackage{balance}       
\usepackage{graphics}      
\usepackage[T1]{fontenc}   
\usepackage{txfonts}
\usepackage{mathptmx}
\usepackage[pdflang={en-US},pdftex]{hyperref}
\usepackage{color}
\usepackage{booktabs}
\usepackage{textcomp}

\usepackage{adjustbox}
\usepackage{booktabs} 
\usepackage{array,multirow}

\usepackage{microtype}        
\usepackage{ccicons}          

\usepackage{todonotes}

\def\plaintitle{SIGCHI Conference Proceedings Format}

\def\emptyauthor{}
\def\plainkeywords{Authors' choice; of terms; separated; by
 semicolons; include commas, within terms only; required.}

\makeatletter
\def\url@leostyle{%
  \@ifundefined{selectfont}{
    \def\UrlFont{\sf}
  }{
    \def\UrlFont{\small\bf\ttfamily}
  }}
\makeatother
\def\pprw{8.5in}
\def\pprh{11in}

\definecolor{linkColor}{RGB}{6,125,233}
\hypersetup{%
  pdftitle={\plaintitle},
  pdfauthor={\emptyauthor},
  pdfkeywords={\plainkeywords},
  pdfdisplaydoctitle=true, 
  bookmarksnumbered,
  pdfstartview={FitH},
  colorlinks,
  citecolor=black,
  filecolor=black,
  linkcolor=black,
  urlcolor=linkColor,
  breaklinks=true,
  hypertexnames=false
}

\begin{document}

\title{Studying Preferences and Concerns about Information Disclosure in Email Notifications}

\numberofauthors{3}
\author{%
  \alignauthor{Yongsung Kim\\
    \affaddr{Northwestern University}\\
    \affaddr{Evanston, IL}\\
    \email{yk@u.northwestern.edu}}\\
  \alignauthor{Adam Fourney\\
    \affaddr{Microsoft Research}\\
    \affaddr{Redmond, WA}\\
    \email{adamfo@microsoft.com}}\\
  \alignauthor{Ece Kamar\\
    \affaddr{Microsoft Research}\\
    \affaddr{Redmond, WA}\\
    \email{eckamar@microsoft.com}}\\
}


\maketitle

\begin{abstract}
The proliferation of network-connected devices and applications has resulted in people receiving dozens, or hundreds, of notifications per day. When people are in the presence of others, each notification poses some risk of accidental information disclosure; onlookers may see notifications appear above the lock screen of a mobile phone, on the periphery of a desktop or laptop display, or projected onscreen during a presentation. In this paper, we quantify the prevalence of these accidental disclosures in the context of email notifications, and we study people's relevant preferences and concerns. Our results are compiled from an exploratory retrospective survey of 131 respondents, and a separate contextual-labeling study in which 169 participants labeled 1,040 meeting-email pairs. We find that, for 53\% of people, \emph{at least} 1 in 10 email notifications poses an information disclosure risk. We also find that the real or perceived severity of these risks depend both on user characteristics and attributes of the meeting or email (e.g. the number of recipients or attendees). We conclude by exploring machine learning algorithms to predict people's comfort levels given an email notification and a context, then we present implications for the design of future contextually-relevant notification systems.
\end{abstract}

%
%


%
%


\maketitle

\section{Introduction}
\begin{figure}[t]
\centering
\includegraphics[width=0.45\textwidth]{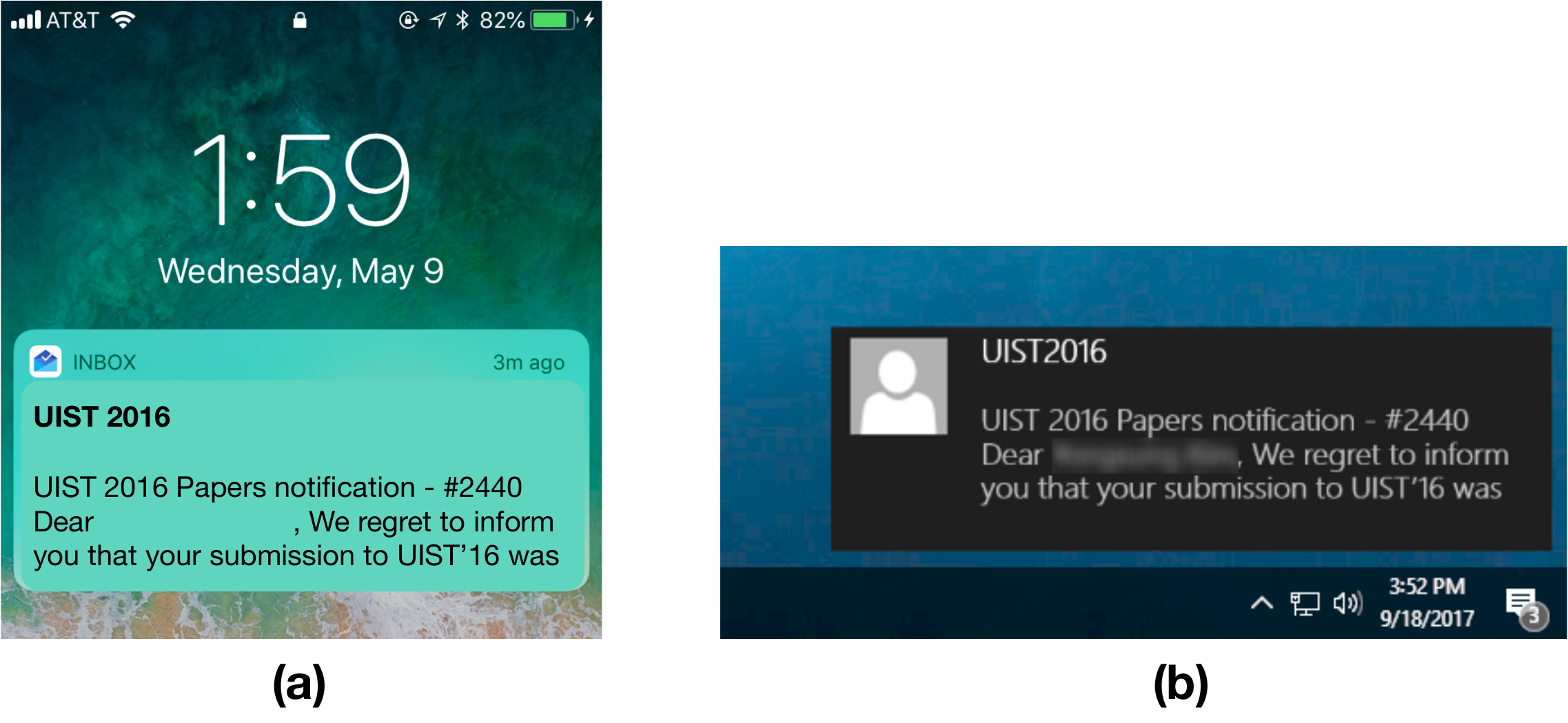}
\caption{An example of an email notification on an iPhone (a) and on a computer running the Windows 10 operating system (b). The notification reveals the email's sender, subject, and first sentence from the message body.
When in the presence of others (e.g, during meetings), users may be uncomfortable receiving notifications and revealing these data to onlookers.}
\label{fig:example_notification}
\end{figure}

Estimates suggest that people receive dozens, or hundreds, of notification messages per day \cite{shirazi_chi14, pielot_mobilehci14, iqbal_cscw10} delivered to a range of connected devices that people carry with them, or that are ever-present in the environment (e.g., wearables, smartphones, computers or -- increasingly -- internet of things devices). Extensive prior research has explored the productivity costs of mal-timed notifications \cite{fogarty_tochi05, iqbal_cscw10, iqbal_chi05}, but little is known about the privacy cost of such messages; people are frequently in the presence of others when notifications arrive, and, in these contexts, each notification poses some risk of accidental information disclosure. For example, email notifications often reveal the sender and subject fields of messages, and may be visible above the lock screen of a mobile phone (Figure \ref{fig:example_notification}a), on the periphery of a co-located desktop or laptop (Figure \ref{fig:example_notification}b), on a projection screen during a presentation, or on a smart TV ~\cite{weber_tvx16}. As Susan Farrell writes in \cite{farrell}:

\begin{quote}
"By making smart devices ubiquitous, we've exposed ourselves to computer-assisted embarrassment. We must expand our usability methods to cover not only the isolated user in one context of use, but also the social user, who interacts with the system in the presence of others."
\end{quote}

To mitigate the potential privacy costs of notifications, current operating systems or applications provide limited notification strategies that offer either binary control (e.g. turning on and off notifications) or minimal information level control (e.g. whether or not to show previews). Using the binary control strategy, users have to make trade-offs between minimizing the privacy risks in accidental information disclosure by suppressing notifications at the cost of losing timely access to potentially important, or time-critical, information. Using the minimal information level control strategy, although users have some control over the level of information that is being disclosed, they are unnecessarily bound to one solution for all notifications regardless of their contexts and the contents of the notifications. For example, if a user is expecting certain emails pertaining to their current meeting, the user may wish to know when those emails arrive (by indicating either sender and/or the preview of the email), but may not want to disclose the sender or the preview of other emails (e.g. shipping details, emails from their doctors).

With the ultimate goal of designing more contextually-relevant notification strategies to provide sufficient information while still preserving user privacy, as a first step, we sought to understand the information disclosure risks arising from notifications received when people are in the presence of others. Specifically, we ask the following research questions:
\begin {itemize}
\item \textbf{RQ1:} How often do email notifications pose an information disclosure risk? I.e., How often are emails received while in the presence of others, and how often do these emails contain sensitive information?
\item \textbf{RQ2:} To what degree are people's preferences and concerns dependent on user characteristics (personal preferences, job role, etc.) versus being context-dependent with respect to the content disclosed by the notification, and the people in the room?
\item \textbf{RQ3:} To what degree can a machine-learned system anticipate information disclosure risks? If such predictions are possible, what features are important? 
\end {itemize}

To answer RQ1 and RQ2, we conducted two studies in an enterprise environment grounded in the scenario of a person receiving an email notification on their notification-capable devices while attending a meeting. We focus on emails because they are the primary source of notifications in this environment: 86\% of professionals cite email as their primary means of communication, and an average 112.5 billion business emails were exchanged per day in 2015 \cite{radicati}. Moreover, email often carries notifications for other services (Twitter, Facebook, Slack), suggesting that our findings are likely to provide insights about notifications generated by other communication platforms.

We first conducted an exploratory retrospective online survey (N=131) to ask respondents to reflect on recent emails and meetings. From this survey we gained an initial understanding about the existence and the prevalence of the accidental information disclosure problem, and we learned the factors affecting subjects' preferences and mitigation strategies currently used in practice. The exploratory study also allowed a broad scope by asking respondents about both personal and work inboxes and contexts, and allowed respondents to provide rich open-ended responses detailing the reasons behind their stated comfortable levels. Findings from this study guided our design decisions for the second study, which provides a deeper, data-centric investigation into information disclosure risks in enterprise settings. For example, it helped us decide which types of features to extract, which survey questions to ask, and which  measures or scales to use.

Based on the insights we gained from the exploratory online survey, we conducted a contextual data collection study (N=169), which employed a custom-built labeling tool, to collect labels and preferences for email-meeting pairs. The labeling tool integrated directly with participants' email and calendaring accounts, allowing the tool to systematically identify emails that were actually received during meetings, focusing labeling on non-hypothetical cases. For each email-meeting pair, the tool automatically extracts features summarizing high-level characteristics of the user, the content of the email message, and the properties of the meeting; and, it does so in a manner that maintains participant anonymity.

Finally, to answer RQ3, we leverage the data collected by the aforementioned contextual labeling tool to explore feature sets and machine learning algorithms to predict information disclosure risks. Here we explore: (a) features that are associated with users (e.g. job title, depth in an organizational chart, the average number of emails received during meetings); (b)  features that describe a particular email message (e.g. number of recipients in an email, categorization of email body); and (c) features that describe a meeting instance, such as whether or not a manager is present, or the number of external members in a meeting.

The remainder of the paper is structured as follows: We review related work, then discuss the findings of the retrospective survey. We describe the second study, which involved the deployment of a custom contextual-labeling tool, then describe relevant findings. Finally, we explore machine learning algorithms to predict people's comfort levels, and we present implications for the design of future social-context aware notification systems. 

\section{Related Work}
\subsection{Productivity Cost in Notifications}
Previous works in notifications mainly focused on the negative impact of interruptions on productivity by quantifying the cost of interruptions and identifying low-cost interruptible moments to send notifications. Previous studies show that the cognitive load of current tasks~\cite{cutrell, iqbal_chi05}, activity transitions ~\cite{ho_chi05, fischer_mobilehci11}, physical activities and user interactions on devices ~\cite{okoshi_ubicomp15}, and other context such as time and location ~\cite{pejovic_ubicomp14} can be used to identify interruptible moments for notifications. In desktop office settings, Fogarty et al. ~\cite{fogarty_tochi05} show how sensor-based statistical models can be used to predict one's interruptibility by leveraging features such as: phone use, ambient noise (e.g., to detect in-person conversations), mouse movement, and keyboard keystrokes. Likewise, COORDINATE ~\cite{coordinate} uses computer activity, calendar information, audio and video signals, and indoor and outdoor location data to predict users' availability. Notification policies then use inferred levels of availability and interruptibility to determine the right time to deliver notifications to users. While these systems demonstrate how systems can avoid interrupting users at inopportune times, research has also recognized that there can be costs associated with delaying notifications. As such, notification policies should trade-off the cost of interruption with the cost of delays  ~\cite{horvitz_cacm03, bayesphone}. 


Our work is distinguished in that we study the privacy cost associated with notifications that disclose some information in social settings, as opposed to most of the previous works focusing on the loss of productivity (e.g. interrupting an important conversation or tasks). The sensitive nature of the problem we study has implications for our methodology for data collection, in which we have taken careful steps to respect respondents' privacy in both study 1 and 2.

\subsection{Privacy Cost in Notifications}

\subsubsection{Privacy Attitudes and Behaviors}
A first step in understanding the potential privacy cost of ill-timed notifications is to understand people's general attitudes towards privacy. Pioneering work in this space includes the Westin Privacy Segmentation Index ~\cite{westin}, that categorizes people into three groups: privacy fundamentalist, marginally concerned, and pragmatist.
\textit{Privacy fundamentalists} are those  who are very concerned about their privacy and very reluctant to share any of their information. \textit{Marginally concerned} are, as the name implies, those who are marginally concerned about their privacy, and are generally willing to share details or data about themselves. \textit{Privacy pragmatists} are people who are somewhat concerned about their privacy, but are willing to compromise some privacy for convenience.

Numerous follow-up studies have adopted these definitions \cite{ackerman_ec99, berendt_cacm05,olson_chi05,acquisti06} to describe general privacy attitudes, but have nonetheless reported finding complex interactions between the types of disclosures, and the audiences who bear witness. For example, \cite{ackerman_ec99} shows that people are generally comfortable sharing their favorite TV shows, favorite snacks, and even email addresses with websites, but are much less comfortable when there is a chance that the website could share the information with others in an identifiable way, or if the information were instead provided by a child under their care. Likewise, Olson et al. show that, even among one's trusted inner circles, one's willingness to disclose information varies greatly depending on the nature of the information being shared~\cite{olson_chi05}. For example, respondents were often uncomfortable disclosing work-related documents with family members, or health information (e.g., pregnancy status) with co-workers.

Adding to this complex motif, past work has also found that privacy attitudes are not always correlated with intended and actual behaviors~\cite{norberg_jca07, spiekermann_ec01, acquisti06}. For example, ~\cite{spiekermann_ec01} found that the participants' actual behavior in an online shopping scenario was different from their self-reported privacy preferences. ~\cite{consolvo_chi05} found that the general privacy attitude was a poor predictor of whether participants would share their location information with other people. Like privacy attitudes, privacy behaviors are context-specific ~\cite{consolvo_chi05, fishbein, woodruff_soups14} and multi-dimensional~\cite{multidimension}.

In this paper we recognize the potential for these complex context-dependent attitudes and behaviors. We designed both of our studies to collect a multitude of signals that characterize how information, audiences, and contexts interact with one another to create situations in which people are uncomfortable (or comfortable) receiving notifications. Additionally, we designed our second study such that it grounds data collection on specific historical instances in which people are known to have received notifications while in the presence of others, rather than asking people about general scenarios and attitudes.

\subsubsection{Difference in Information Disclosure and Privacy Management Strategies}
In addition to understanding privacy attitudes in a broader context, it is also important to understand how notifications may differ from the scenarios explored in past work. In particular, much of the prior work has examined intentional sharing of information with websites or social networking services (SNSs) ~\cite{torabi_soups16, acquisti06, berendt_cacm05, vitak_cscw14}. Granted that when people disclose their information to these services they don't have perfect knowledge of the consequences, with few exceptions, they have agency in deciding which information to disclose. In particular, people often perform an informal risk-benefit analysis when deciding to take action that may have privacy implications ~\cite{dourish_hci06, li_dss12, dupree_chi16, woodruff_soups14}. People also use this agency to enact preventive strategies, including self-censoring, managing access control groups, and taking actions to conceal their identity ~\cite{lampinen_chi11, vitak_cscw14}. In the context of `shoulder surfing'~\cite{eiband_chi17}, recent work also introduced other preventive strategies that can help protect user privacy by detecting onlookers and providing awareness through alerts ~\cite{zhou_chi16}.


In contrast, this level of agency and control is not available in cases where people receive push notifications. Instead, the push notification scenario bears some resemblance to the scenario in which people are tagged in photos shared to SNSs without consent. In such cases, the users of SNSs can often take corrective actions by either untagging the photos or deleting the contents before more people see the posts ~\cite{lampinen_chi11}. With push notifications, disclosures are instantaneous, and it is unclear what corrective actions can be taken after the event.

Most closely related to our work on accidental information disclosure in notifications is accidental information disclosure in web browsing activities in the presence of other people~\cite{hawkey_chi06} and receiving notifications while watching videos together on a smart TV~\cite{weber_tvx16}. In ~\cite{hawkey_chi06}, through a survey of 155 participants, and with three distinct hypothetical web browsing scenarios (i.e., embarrassing, neutral, positive), researchers assessed participants' comfort level in the presence of different groups of viewers. People's comfort level is higher when the viewers are spouses or close friends, and lower with colleagues and supervisors. Also, the comfort level is related to the level of control the participants have (e.g., whether the participants are having control over the mouse and keyboard). In ~\cite{weber_tvx16}, through a survey of 167 participants, researchers assessed the comfort level in different notification variants when people are watching videos on a smart TV with others. The results show that people's comfort level is higher when the notification reveals less information (e.g. notification indicators), and the comfort level decreases as the notification reveals more information. In this work, we consider a richer set of contexts (e.g. meeting type or location, social structure of people in the room, number of people in the room, etc) that may be indicative of comfort level in accidental information disclosure.

Common themes to the above-mentioned research are that one's level of agency/control and the available mitigation strategies, in addition to audience and information types, strongly influence people's comfort levels. As a part of our exploratory study (Study 1), we sought to extend our understanding of the strategies that people employ to mitigate the risks of accidental information disclosure arising from push notifications. In the next section, we briefly review some of the technological controls currently available in the market.


\subsubsection{Limitations and Trade-offs in Existing Notification Strategies}
Existing mobile devices and applications' notification strategies are limited in that they allow users to manage notifications either with binary control (e.g. turning on and off notifications) or minimal information level control (e.g. whether or not to show previews). Although binary control strategies enable users to turn off notifications entirely to avoid accidental information disclosure, users miss the convenience and benefits of receiving information via notifications, instead requiring user to continuously monitor their applications for new information. This means that users have to make trade-offs between minimizing the privacy risks in accidental information disclosure by suppressing notifications at the cost of losing timely access to potentially important, or time-critical, information. Minimal information level control strategies give users some control over information being revealed in the notifications either showing the previews or just a notification indicator (e.g., ``You have a new message'')\footnote{Some applications (e.g., WhatsApp) and mobile devices (e.g., iPhone X) started using notification indicators as a default notification setting.}, but it ineffectively enacts one solution to all notifications without taking into consideration user contexts and notification contents.


As we have more and more internet-connected devices and services that are capable of sending timely, useful and relevant information via notifications, there is a need to better understand the preferences and concerns about accidental information disclosures resulting from these notifications. Likewise, there is a need to explore designs and notification strategies that can adapt to the user's context, as well as to the content of the notifications.

\section{Study \#1: Exploratory Retrospective Survey}
As a first step, we describe the results of an exploratory retrospective survey which was designed to gain our initial understanding of the following: (1) how often notifications pose an information disclosure risk? (2) How do features of the notifications, meetings and individuals contribute to the perceived risk? And, (3) what mitigation strategies do people currently employ. As noted in the introduction, the methods employed here allow a broad investigation covering numerous contexts (e.g., personal vs. work, email topics and themes, etc.)


\subsection{Procedure}
The survey began by collecting basic demographic information including education, job role, age, gender, notification-capable device use, average number of meetings in a day. It then asked participants to answer questions about a notification scenario, which was evolved slowly over several sections of the questionnaire, as follows:

\begin{enumerate}
\item Respondents were first asked to consider their most recent meeting. Respondents answered questions about their role in the meeting, their relation to each of the meeting's attendees, and the meeting's time and location, and the types of notification-capable devices that were present (including Desktop PC, Laptop, Smartphone, Smartwatch, and Smart speaker).


\item The survey then asked respondents to open their primary work email inbox, consider their 10 most recent emails (excluding the survey invitation), and imagine a scenario in which they received notifications for those emails during the aforementioned meeting. Respondents were asked about the age of the 10\textsuperscript{th} email.\footnote{We decided to only rely on a respondent's 10 most recent emails to (a) convey a simple and consistent sampling criteria in the retrospective survey, and (b) minimize response and counting errors by allowing respondents to observe their emails within a single window pane (the decision made after pilot testing). To partially compensate for this limitation, we additionally asked the age of 10\textsuperscript{th} email, which enables us to calculate the incoming email rate.} Furthermore, we asked participants to imagine that their notification-capable devices' screens were visible to meeting attendees, and answer how many email notifications (out of the ten emails) they would have been uncomfortable sharing with the people in the room. We refer to this scenario as a \emph{hypothetical information disclosure event}, or \emph{HIDE}.

\item The questions above were then repeated for the 10 most recent emails in respondents' personal email inboxes.

\item Respondents were then asked to select one email (could be from either work or personal email inbox), if applicable, that they would have been most uncomfortable receiving in the scenario outlined above. We refer to this email as the \emph{HIDE email}. Respondents were asked to rate their comfort level in sharing the HIDE email notification with the people in the room with a 5-point Likert scale (1: Very uncomfortable, 5: Very comfortable). Respondents then were asked to describe general features of the HIDE email, including their relation to its sender, the number of recipients, the type of the inbox (work or personal), the type of content contained therein, and to explain, in broad terms, why receiving notifications for this email would be uncomfortable.


\item Finally, respondents were asked to detail the mitigation strategies they employ to minimize the information disclosure risks presented by email notifications.
\end {enumerate}

The survey was deployed by emailing a random sample of 800 employees within a large IT corporation. There were 21 email delivery failures, for various technical reasons, resulting in 779 individuals successfully receiving the invitation. We describe our findings next.



\subsection{Results}
\subsubsection{Participants}
We received 118 completed, and 13 partially completed responses (response rate = 17\%, completion rate = 90\%). Of the 131 total respondents, 85 were male (65\%), 44 were female (34\%), 2 preferred not to answer. Ages were distributed as follows: 18 -- 24 years old (5\%), 25 -- 34 (24\%), 35 -- 44 (31\%), 45 -- 54 (27\%), 55 -- 64 (8\%), $\ge65$ (2\%), and 2\% declined to answer.

Participants reported occupying a diverse set of job roles including: program managers (28\%), software developers (19\%), marketing and sales people (8\%), and IT support staff (8\%). The remaining 49 individuals (37\%) worked in diverse roles such as administrative assistants, data scientists, designers, attorneys or other roles in the legal department, and human resources.

\subsubsection{Prevalence of the information disclosure risk}
One key finding from the retrospective survey, and a partial answer to our first research question, is that the majority of respondents (53.4\%) reported that \emph{at least one} of their ten most recent work emails would have resulted in an uncomfortable disclosure of information (i.e. they selected a comfort rating of < 3 on the 5-point Likert scale for sharing the sender, subject, and first sentence of the email) had the email arrived during their most recent meeting. This increases significantly to 73.3\% when respondents were asked to consider the ten most recent emails delivered to their personal accounts (two tailed difference of proportions test, $p << 0.001$, $Z = 3.33$). While these findings demonstrate the potential for risk, we cannot be sure how many emails were actually received during the meetings. We address this limitation in the second study.


\begin{figure}[t]
\centering
\includegraphics[width=0.45\textwidth]{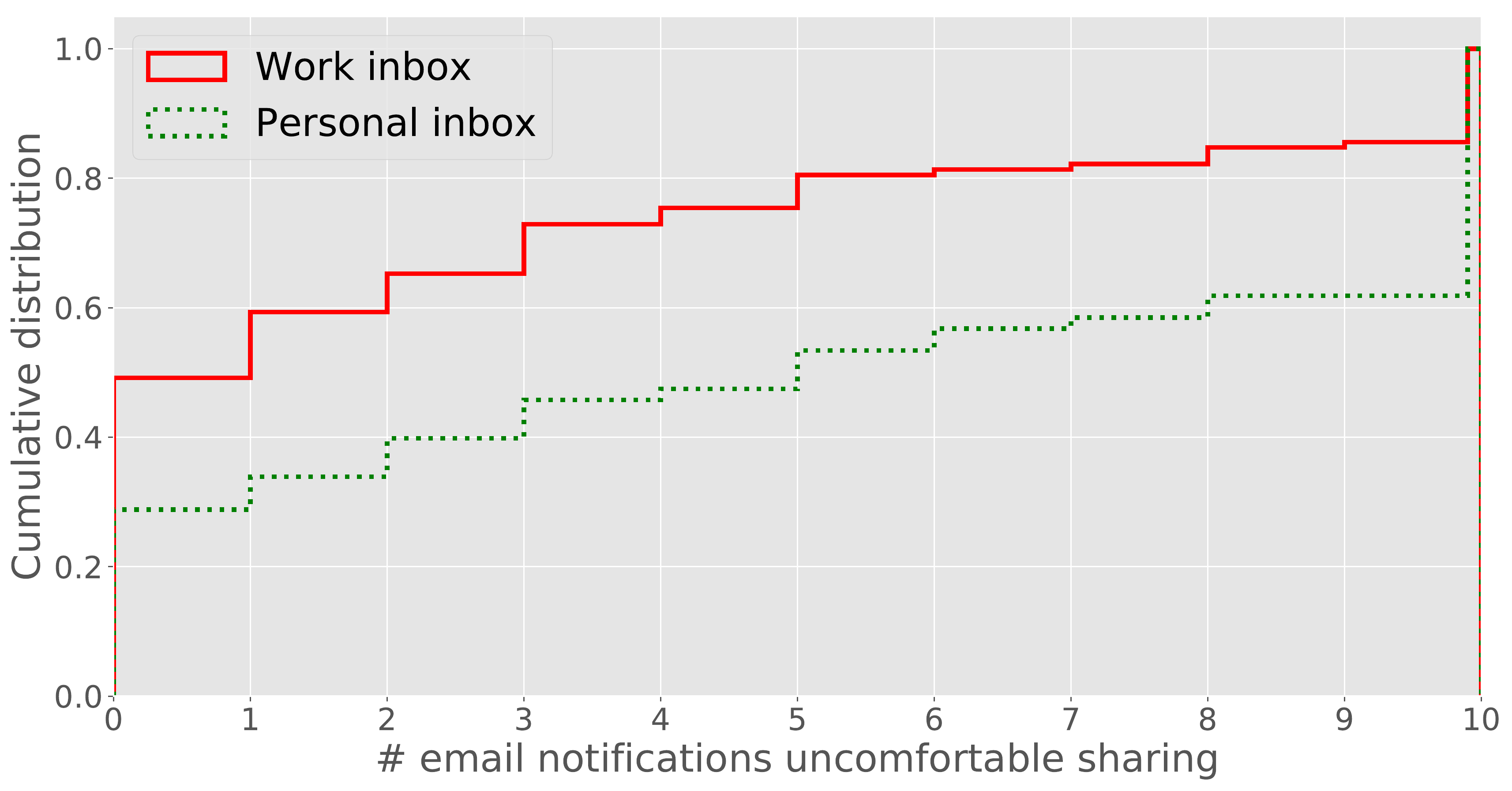}
\caption{Cumulative distribution of the number of work (red, solid) and personal (green, dotted) emails that people felt uncomfortable sharing with meeting attendees.}
\label{fig:work_personal}
\end{figure}

\subsubsection{Three groups of respondents}
Figure \ref{fig:work_personal} extends the above analysis by presenting the cumulative distribution for work emails (red) and for personal emails (green) that participants indicated they would be uncomfortable receiving in their most recent meeting. There are two notable features of these distributions: first, there is a sharp rise at $x = 10$ emails, with 17\% of respondents noting they would be uncomfortable receiving \emph{any} work email notification in the presence of others. This increases to 40\% for personal emails. We also note that the $y$-intercepts of the two curves are rather high: 47\% of respondents reported that \emph{none} of their ten most recent work emails were sensitive. This decreases to 25\% for personal emails. The remaining 36\% and 35\% of respondents were more selective, and reported that \emph{some} of their work and personal emails were sensitive, respectively.



On the surface, these three groups bear some resemblance to Westin's three categories, but we are sensitive to the fact that privacy concerns are complex and multidimensional. The responses we collected may be dependant on user characteristics (e.g. personal preferences, the nature of their occupation), contents in the notifications, and the contexts and their relationship with the people in the meeting. For example, one participant explained that she was uncomfortable sharing the information in notifications due to the nature of her occupation:
\begin{quote}
``\emph{I deal with a lot of privileged and confidential information on an hourly, daily basis. I am not able to share the information and it should not be visible to others.}''--P122
\end{quote}

We could also see that some of the respondents are less concerned about the email notifications divulging information because they personally are less concerned about the contents of notifications being shared with the other people in the room. As P81 reported:

\begin{quote}
``\emph{...But I'm generally not embarrassed by who and what I am. Plus, humor is a good way to diffuse why I get the spam I get.}''
\end{quote}

Also, worth noting that the floor and ceiling effects we observe may simply reflect limitations in our ability to sample the most/least sensitive contexts, contents, and emails for some users. Our second study will use more sophisticated sampling method to address such limitations.



To sum up, the distribution of responses indicate the three distinct groups of respondents.  While some of the open-ended survey responses -- like the ones above --  highlight the relationship between respondents' occupation, personality and demographics~\cite{sheehan_is02} on notification preferences, some indicate that their preferences depend on the context. To understand this relationship, in the next sections, we extend the analysis by examining the content disclosed by notifications, and the people attending the meeting.


\subsubsection{Notification fields and their content}
Given prior research \cite{olson_chi05}, we suspect that one's comfort level depends both on the types of information disclosed by notifications, and on the audience witnessing the disclosure. Addressing the former, participants were asked to select one email that they would have been uncomfortable sharing in their most recent meeting from among their 10 most recent emails. As noted earlier, we refer to this as a hypothetical information disclosure event (HIDE). Participants were asked to answer questions about their HIDE emails, and to describe, in their own words, the types of information that rendered a notification sensitive. We limit the remaining discussion to the 62 respondents (47\%) who both identified a HIDE email, and who fully completed this portion of the survey questionnaire. We describe the general properties of HIDE emails, then analyze open-ended responses.





In most cases (72.6\%), respondents elected to describe HIDE emails that were delivered to their work email inboxes -- perhaps reflecting that the survey invitation was sent during business hours. Among these emails, most were sent by work colleagues (75.6\%), including: direct superiors (15.6\%), team members (28.9\%), and other members of the organization (28.9\%). Conversely, external senders included: clients or customers (8.9\%), family and friends (4.4\%), and one instance each from a doctor, and from an insurance company. Conversely, when HIDE emails were delivered to personal inboxes, most were from friends and family (76.5\%), but also included messages from: doctors, banking institutions and external recruiters.

In the majority of cases (62.9\%) respondents were the only recipient of their HIDE emails -- though there  were 11 cases (12.9\%) where the email was sent to 5 or more people (via the to: or cc: lines). This finding agrees with the intuition that emails sent only to one person are, perhaps, more likely to contain private information. We explore this hypothesis further in the second part of this paper.


As noted above, we asked participants to describe, in broad terms, what types of information rendered these emails sensitive. Open-ended responses were analyzed, and themes identified, following open coding practices \cite{Lazar_2010}. 
Among the top themes were: unspecified personal life details (14 instances), unspecified confidential work documents (11 instances), details of ongoing projects (10 instances), health information (7 instances), personal successes or failures (7 instances), financial communications (4 instances), and messages that mention people attending the meeting (3 instances). These themes largely overlap those identified by Olson et al. in \cite{olson_chi05}.


\begin{figure}[t]
\centering
\includegraphics[width=0.45\textwidth]{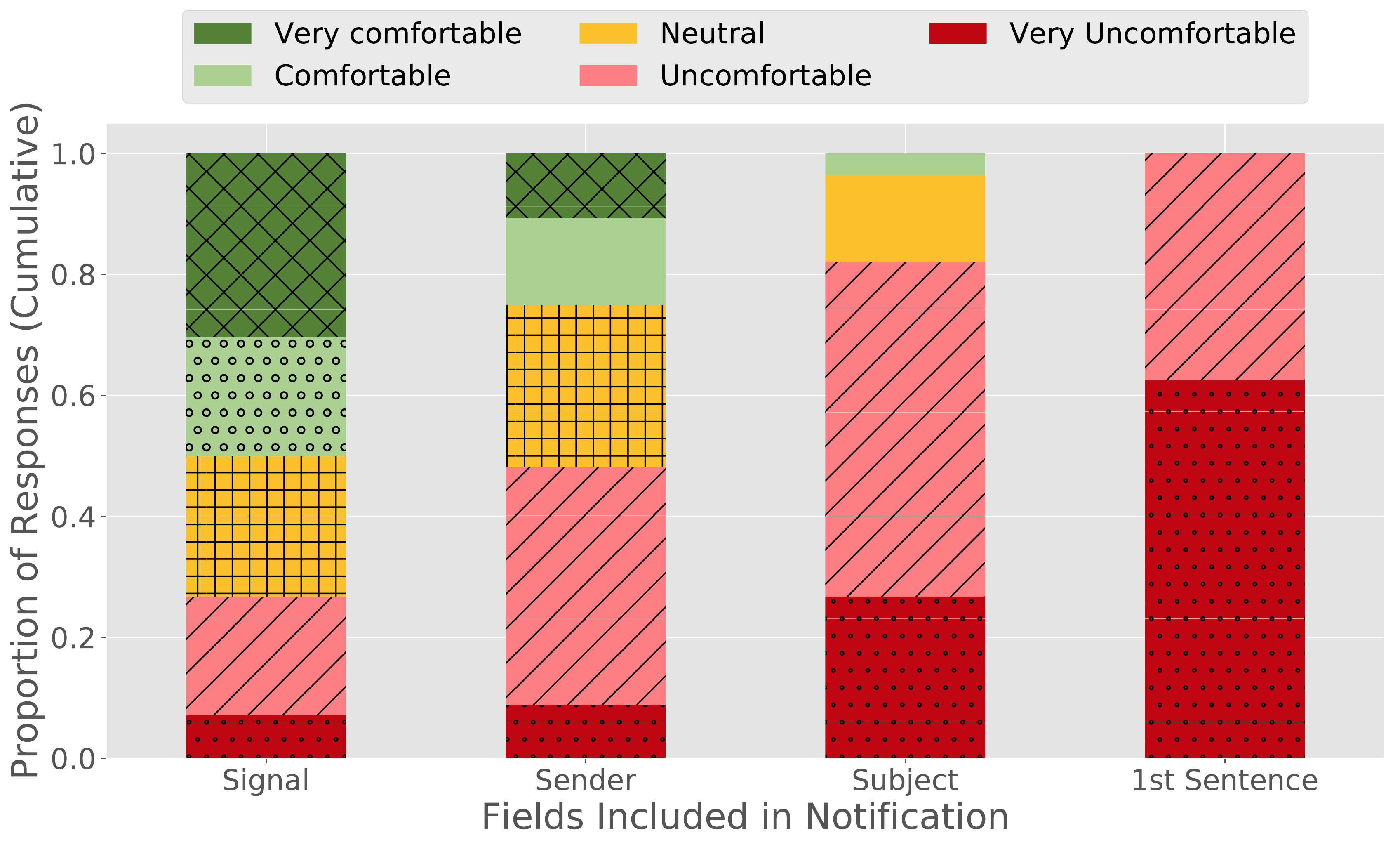}
\caption{Distribution of respondent comfort levels for sharing different fields included in typical email notifications for HIDE emails.}
\label{fig:survey_fields}
\end{figure}

%


As email notifications can reveal numerous email fields (Figure \ref{fig:example_notification}) including: sender, subject, and the first sentence of the email body, we asked participants to rate how comfortable would they have been if the people in the room saw the different email fields. Each field-type reveals a different class of information, and, to varying degrees, poses an information disclosure risk. For example, the sender field reveals that a respondent is in correspondence with a particular individual, while the subject reveals the topic of discussion. Correspondingly, among HIDE emails 25\% of survey respondents were comfortable with email notifications that revealed the sender's identity, while only 3.6\% were comfortable with notifications that reveal the email subject. This difference is highly statistically significant ($p = 0.001$, $Z = 3.20$). Figure \ref{fig:survey_fields} provides further details, breaking down respondent comfort levels by field-type.


\subsubsection{Audience features}
We also explored whether meeting properties (e.g., location, attendance, etc.) impacted the proportion of respondents who identified a HIDE email from among their 10 most recent messages. Though this analysis revealed no statistically significant results, we present our findings to: (1) characterize the meetings our respondents attend, (2) offer points of comparison with the results from our second study, and (3) to identify weak trends that may yet become features in machine-learned models.

The majority of respondents reported that their most recent meeting occurred in a conference room (52.7\%), though people also reported hosting meetings in their own offices (22.9\%), attending meetings in someone else's office (15.3\%), and attending meetings in other common spaces such as a lounge or atrium (8.4\%). Among respondents whose meetings occurred in conference rooms, 50.7\% were able to identify an email they would have been uncomfortable receiving in that context. Likewise, among respondents meeting in offices, 46.6\% were able to identify such an email. These differences are not statistically significant ($p = 0.61$, $Z = 0.509$).


\begin{figure}[t]
\centering
\includegraphics[width=1\columnwidth]{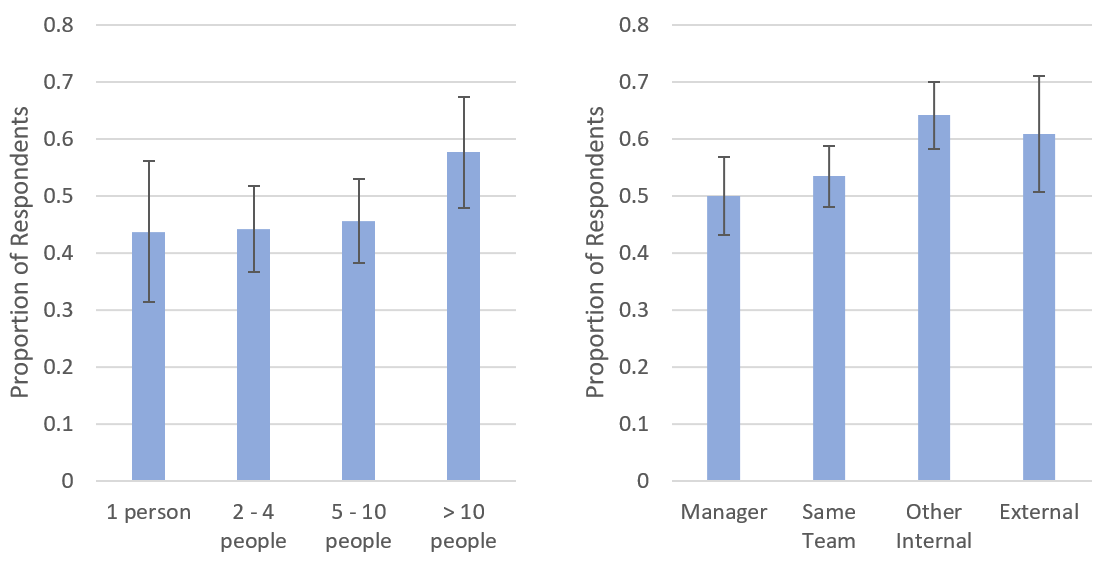}
\caption{Proportion of respondents reporting HIDE emails, partitioned by the number of people attending the meeting (left) and the attendees' relationships to the respondent (right). Error bars show standard error; none of the pairwise differences are statistically significant at the $\alpha = 0.05$ level.}
\label{fig:meeting_size}
\end{figure}

When asked how many people attended the respondent's most recent meeting, the mode response was ``5 -- 10 other people'' (35.1\%). Other response categories included: ``2-4 other people'' (32.8\%), ``10 or more people'' (14.5\%), and ``1:1 meetings'' (12.2\%). Again, we examine the proportion of respondents who were able to identify an email they would have been uncomfortable receiving in each of these contexts. This proportion monotonically increases with the size of the meeting (Figure \ref{fig:meeting_size}, left), but the pairwise differences are not significant.






Finally, we report that meetings were attended by team members (65.6\%), direct superiors (41.2\%), other members of the respondent's organization (51.1\%), as well as people external to the organization (17.6\%). In each case, we examine the proportion of respondents who were able to identify an email they would have been uncomfortable receiving in the meeting. Figure \ref{fig:meeting_size} (right) shows that the proportion increases as the meeting's attendee list grows beyond one's own team. Though this agrees with intuition, the differences are not statistically significant. We revisit this observation later in this paper.

\begin{figure*}[t]
\centering
\includegraphics[width=0.9\textwidth]{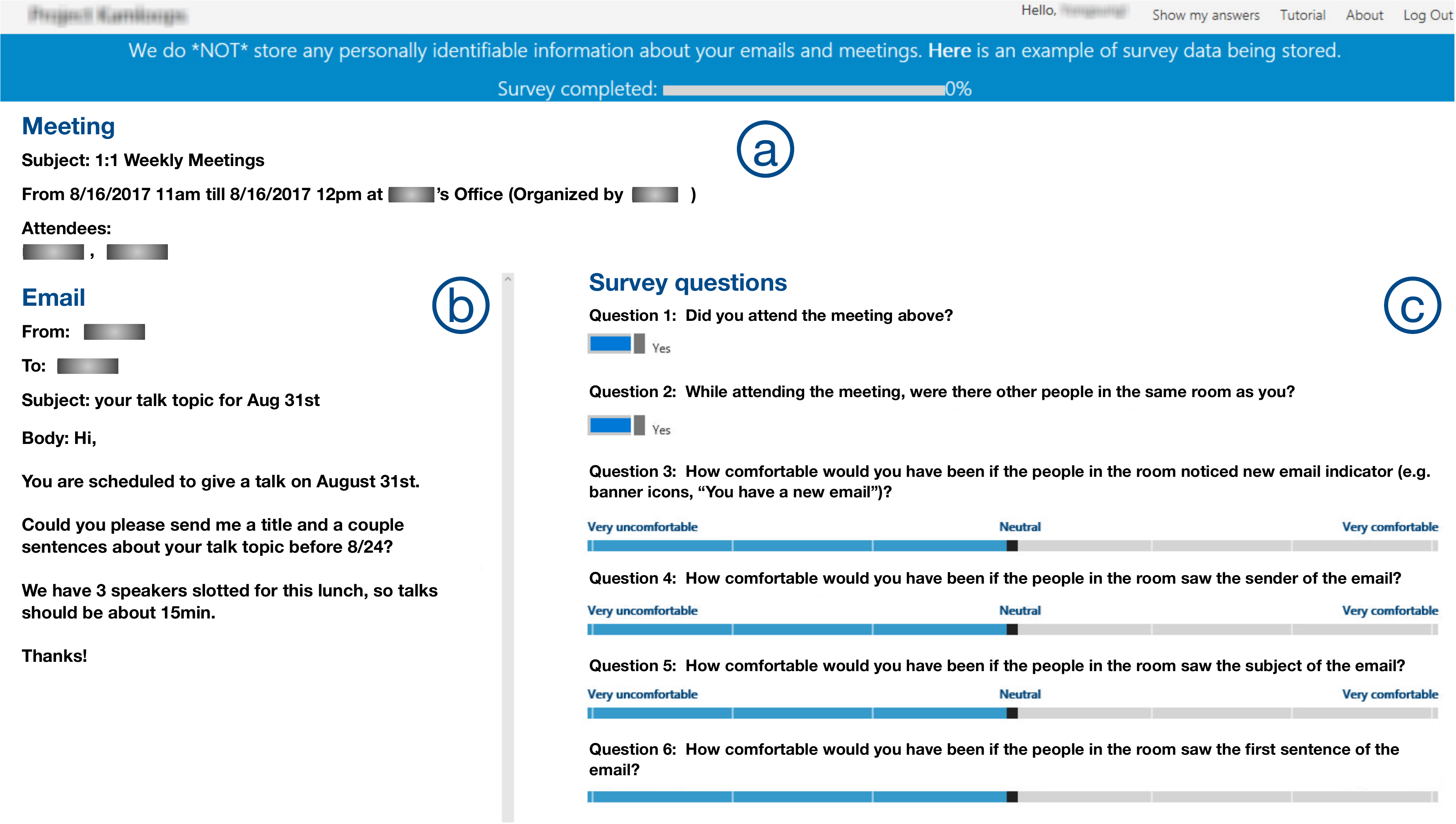}
\caption{Contextual labeling interface for survey questions about various email-meeting pairs. (a) Meeting pane that shows a randomly chosen meeting, time and location of the meeting, attendees, and subject of the meeting; (b) Email pane that shows a randomly chosen email that was received during the meeting, the sender, recipients, subject, and body of the email; (c) survey questions about their comfort level in sharing different fields of the email, type of the email, preferences for different devices and disclosure level.}
\label{fig:interface}
\end{figure*}

\subsubsection{Current mitigation strategies}
We asked respondents to detail any mitigation strategies they employ on mobile devices, as well as on laptop or desktop computers, to minimize the information disclosure risks associated with ill-timed email notifications.

On desktop or laptop computers, the most common mitigation strategy was to disable all notifications in software, e.g., by closing applications (67 instances), or by explicitly enabling the ``do not disturb'' feature (53 instances). This highlights the limitation of binary control strategies in suppressing notifications for user privacy at the cost of losing timely access to potentially important information. Conversely, on mobile phones respondents were most likely to report physical mitigation strategies such as keeping their phones in a pocket (63 instances), or flipping the device face down so that its screen was not visible to the people in the room (53 instances). These physical mitigation strategies mimic minimal information level control strategies as an audio or a vibrotactile cue will be served as a notification indicator and notifications would continue to be perceived by the respondent.






In summary, results from the exploratory retrospective survey provide a key set of initial insights about the prevalence of the information disclosure risks posed by notifications, and about how email topics and fields may contribute to this risk. Moreover, we found that respondents could be clustered into three distinct groups based on the proportion of emails they received that would result in uncomfortable notifications.

While the survey was designed to ground responses in respondents' actual emails and meetings, it is limited in two important ways. First, to convey a simple and consistent sampling criteria and to minimize response or counting errors, the survey asked respondents to comment only on a single meeting, and on their single most-sensitive email. Second, it asked respondents to consider \emph{hypothetical} situations in which they receive the most-sensitive email during the most recent meeting. Together, these survey design choices limit our ability to measure actual incidence rates, and to model the overall distribution of risk. To address these limitations, we developed and deployed a custom-built contextual labeling tool, which we describe in the next section.

\section{Study \#2: Contextual Labeling Study}
Based on the insights we gained from the exploratory retrospective study, we developed and deployed a tool (Figure \ref{fig:interface}) to extract various features and collect labelled data for learning a context-dependent predictive model of disclosure risk. The tool allows participants to: view emails that they received during meetings, view the details of those meetings, and rate their comfort-levels in receiving the corresponding email notification in that context. We described the procedure, apparatus, and results below.


\subsection{Procedure and Apparatus}
The contextual-labeling study was deployed within the same large information technology company as the initial retrospective survey. Participants shared a common computing environment. In particular, they stored their emails and calendars in a common email and calendar web service. This homogeneous environment greatly simplified the administration of the study, and the implementation of the tool.

\subsubsection{Procedure}
Participants were recruited by emailing a random sample of 4000 employees, distinct from the 800 who were contacted for the retrospective survey. The invitation email described the study's purpose and procedure, and provided sufficient information to validate the authenticity of the invitation (e.g., links to internal systems and pages documenting the experiment, and the results of both an internal review process, and IRB review). Crucially, the invitation also included a link to the web application that hosted the labeling tool.

Upon navigating to the web application, participants were first shown the purpose of the study (i.e. characterizing people's preferences about information disclosed by desktop and mobile notifications that arrive when the intended recipient is in the presence of others). Then, participants were asked to authenticate to the tool using their corporate credentials, and to grant the tool time-limited access to their corporate email and calendaring accounts. Once participants were authenticated, they were presented with a brief tutorial of the labeling tool, and its three regions:

\begin{itemize}
\item The top region (Figure \ref{fig:interface}a) displayed a recent meeting, randomly selected from a day.
Visible fields included: meeting subject, time, location, organizer, and a list of attendees.\\
\item The left region (Figure \ref{fig:interface}b), displayed a randomly selected email that arrived during the meeting. Visible fields included: the sender, recipient list (the `to:' and `cc:' lines), subject, and email body.\\
\item The right region (Figure \ref{fig:interface}c) contained a short survey, where participants could provide labels and answer questions about the email-meeting pair. Questions asked about participants' comfort levels in having notifications disclose various fields of the email to the people attending the meeting.
\end{itemize}

To collect data across a wide range of email-meeting pairs, the tool samples one email-meeting pair per day, moving backward in history one day at a time until 10 pairs are labeled. In each day, a meeting was randomly selected and then an email that was received during the meeting was randomly selected as well. If participants indicated that they did not attend a scheduled meeting, or that the meeting was conducted via teleconference, the tool selected another email-meeting pair for labeling.

Upon inputting labels for 10 email-meeting pairs, participants were presented with a debriefing page,  
and an optional invitation to take part in a raffle for one of three \$50 Amazon.com gift cards. This sweepstakes was conducted in appreciation for their participation.

In addition to collecting user-provided labels and preferences, the labeling tool collected high-level features of the emails and meetings (Table \ref{table:features}).  Importantly, the study was conducted completely anonymously: users were assigned random session identifiers, the participation sweepstakes was conducted on a separate unconnected system, and the features were chosen to be non-personally identifiable. We provide more details about these features in the next section.

\begin{table}
\centering

\adjustbox{width=0.45\textwidth}{\begin{tabular}{|c|p{3cm}|p{7cm}|}
 \hline
 & \textbf{Feature} & \textbf{Description} \\
 \hline
 \multirow{6}{*}{\rotatebox[origin=c]{90}{User}} & jobTitle & k-anonymized job title\\
 & orgDepth & depth in the organizational chart\\
 & numEmails & number of emails received in the past week\\
 & numMeetings & number of meetings scheduled in the past week\\
 & avgEmlPerMtg & average number of emails received in meetings\\
 & numMtgWithEml & number of meetings interrupted by emails\\
 \hline
 \multirow{12}{*}{\rotatebox[origin=c]{90}{Email}}  & numRecipients & number of recipients in the email\\
 & numDistList & number of distribution lists as the recipient\\
 & numThreads & number of threads in the email\\
 & isAutogenerated & is autogenerated email\\
 & numPplMentioned & number of attendees mentioned in the email\\
 & numAttachment & number of attachments\\
 & attachment[type] & type of attachment in the email\\
 & isSenderInternal & is the sender internal to the organization\\
 & numSubjectWords & number of words in the email subject\\
 & numBodyWords & number of words in the email body \\
 & entity[type] & mentions of people, locations and organizations\\
 & LIWC[cat] & feature vector over email body\\
 \hline
 \multirow{10}{*}{\rotatebox[origin=c]{90}{Meeting}} & location & meeting location\\
 & numAttendees & number of people attending the meeting \\
 & isManagerPresent & is the person's manager present\\
 & numDirectReports & number of direct reports present\\
 & numOrgAbove & number of attendees above in the org chart\\
 & numOrgBelow & number of attendees blow in the org chart\\
 & numExternal &  number of attendees external to organization\\
 & numSubjectWords & number of words in the meeting subject \\
 & numBodyWords & number of words in the meeting body \\
 & LIWC[cat] & feature vector over meeting body\\
 \hline
 \end{tabular}}

\caption{A list of user, email and meeting features that the labeling tool automatically computed for each email-meeting pair labeled by participants.}
\label{table:features}
\end{table}

\subsection{Feature Extraction}
Table \ref{table:features} presents a list of features automatically collected by the labeling tool. Features are broadly categorized into three groups. User features are those that are associated with the participant, but not with a particular email-meeting pair. For example, this class includes the average number of emails a person received during meetings in the past 7 days, their depth in the organization chart, and their job title. To preserve anonymity, we used k-anonymization ($k=50$) at the organizational level for full job titles, and separately, for job-title bigrams. Email features are those that describe a particular email message. Examples of these features include the number of recipients, Linguistic Inquiry and Word Count (LIWC, \cite{tausczik_2010}) feature vectors, mentions of people, locations, and organizations, and whether the email is machine-generated. For LIWC categories, we tokenized the body, categorized each token into the pre-defined psychological and linguistic categories in LIWC, then computed the percentage of tokens in each category relative to the email body as a whole. We also used the Stanford named entity recognizer \cite{Finkel_2005} to detect mentions of people, places and organizations. To check whether or not an email is autogenerated, we used a simple heuristic to check whether or not an email contains `Unsubscribe.' Meeting features describe the meeting instance, and include details such as the number of attendees, and whether a person's manager is in attendance. Organizational relationships were computed by cross referencing the attendee information from the calendar and the company's organizational chart as follows: for each attendee, we checked whether or not the attendee was the person's manager, direct report, above or below the org chart. Finally, we include one hybrid feature which counts the number of meeting attendees mentioned in the email.

\begin{table*}[t] \centering
\begin{tabular}{@{\extracolsep{1pt}}lccc}
\\[-1.8ex]\hline
\hline \\[-1.8ex]
 & \multicolumn{1}{c}{10 Uncomfortable} & \multicolumn{1}{c}{Mixed Comfortable} & \multicolumn{1}{c}{10 Comfortable} \\
\hline \\[-1.8ex]
Labeled meeting-email pairs & 23\% & 36\% & 41\% \\
Retrospective survey (work) & 17\% & 36\% & 47\% \\
Retrospective survey (personal) & 40\% & 35\% & 25\% \\
\end{tabular}
\caption{Distribution of three groups of respondents in both the retrospective and contextual-labeling study. ``10 Uncomfortable'' indicates respondents who labelled all 10 emails as uncomfortable, ``10 Comfortable'' indicates respondents who labelled all 10 emails as comfortable, and ``Mixed comfortable'' indicates respondents who reported a mixture of both comfortable and uncomfortable emails.}
\label{tab:user_classes}
\end{table*}

\subsection{Results}
We sought to replicate the analyses we conducted in the exploratory retrospective survey, when possible. As we will show, the consistency of their results helps bolster our confidence in their validity and reliability. Now, we discuss our main findings of this study.

\subsubsection{Participants}
In total, we received 1,040 meeting-email pairs labeled by 169 participants. Similar to the retrospective survey, job roles were diverse. The largest two categories included software developers (21.3\%) and program managers (13.6\%). An additional 52 individuals (30.8\%) occupied various roles including: marketing managers, attorneys, sales specialists, business planners, etc. Finally, the job roles of 58 participants (34.3\%) were filtered by k-anonymization.


%
%

%

In addition to collecting job demographics, the labeling tool collected general measurements of the participants' calendars and email inboxes. For performance reasons, these coarse measurements are constrained to examining users' 300 most recent emails, together with the last 7 days of their calendar. 75.15\% of participants' inboxes contained fewer than 300 emails received during this 7-day window (average: 103.7), allowing a full measurement of email-meeting co-occurrences during the week. These participants' 7-day calendars contained an average of 13.84 meetings (SD: 9.04). On average participants received emails during 7.35 (53\%) of these meetings  (SD: 6.10). In other words, slightly more than half of all meetings were interrupted by email.

A similar analysis can be performed for the remaining 24.85\% who received more than 300 emails. Here, the results cover a variable time frame that is necessarily shorter than a week. For these participants, their most recent 300 inbox emails co-occurred with an average of 8.74 meetings (SD: 6.14). These participants' 7-days calendars contained an average of 18.83 meetings (SD: 11.66), suggesting that \emph{at least} 46\% of their meetings are interrupted by email.

The remaining analysis considers specific email-meeting pairs that are sampled from participant's calendars and email inboxes. Unlike the above-mentioned aggregate measures, the sampling procedure is not constrained by the 7-day, 300-email, limit.


\subsubsection{Prevalence of the information disclosure risk}
As detailed in the procedure section, participants were asked to answer questions about 10 email-meeting pairs. To generate these pairs, we randomly sampled a meeting, then randomly sampled an email received during the meeting. Mirroring the analysis of the retrospective survey, 90 out of 169 people (53.3\%) had at least one email whose notification they rated as uncomfortable sharing (i.e., they selected a comfort rating of $<4$ on the 7-point Likert scale for sharing the sender, subject and first sentence of the email). This proportion is nearly identical to that which was found in the retrospective survey for emails delivered to work inboxes (53.4\%, Figure \ref{fig:work_personal}).


\subsubsection{Three groups of respondents}
Earlier, when we analyzed the results of the retrospective survey, we observed that respondents could be bundled into three groups based on how they rated their comfort levels in sharing the notifications of their 10 most-recent emails with the attendees of their most recent meeting. Now, in this labeling study, participants were asked to label 10 emails known to have actually arrived during 10 distinct meetings. This allows a more ecologically valid analysis of this phenomenon. Results are presented in Table \ref{tab:user_classes}, together with the distribution reported in our retrospective survey. The results show that 23\% of the respondents were uncomfortable sharing any of 10 emails in the respective meetings (indicated under the heading ``10 Uncomfortable'' in Table \ref{tab:user_classes}), while 41\% of the respondents were comfortable sharing all of the 10 emails (10 Comfortable in Table \ref{tab:user_classes}), and 36\% of them were uncomfortable sharing some emails (``Mixed Comfortable'' in Table \ref{tab:user_classes}). Notably, the distribution of users is roughly consistent across both the retrospective survey and the labeling tool (for work inboxes). This suggests that, although the retrospective survey involved a hypothetical situation, its findings closely match those of real-world scenarios. The distribution of the groups 
also resembles the classical trichotomy of the Westin Index; but, again, we are sensitive to the fact that these preferences can be highly nuanced and context-sensitive. To that end, we explore contextual factors below, as well as later in the section entitled ``Predicting Comfort Level''. 

\subsubsection{Email and meeting properties}
When reporting the results of the retrospective survey, we examined various properties of emails and meetings that might indicate, or themselves constitute, an information disclosure risk. We now reexamine those criteria with data compiled from this study.

\textbf{Number of email recipients:} In the retrospective survey, we found that the number of people in an email's recipient list may influence users' comfort levels in receiving notifications. Specifically, we observed that the majority of HIDE emails (62.9\%) had only a single recipient. We can report a similar statistic for data collected via the labeling tool but must first filter the data such that they are directly comparable: In the retrospective survey, users were asked to discuss the most sensitive email from among those they would be uncomfortable sharing in a meeting. When we apply the same criteria to label-tool data, we find that 46.8\% of emails have only a single recipient. 

The labeled data also allows for a more deliberate examination of this phenomenon. This is because it contains examples of both sensitive and non-sensitive email-meeting pairs. Over all 1040 pairs, 406 (39\%) emails were delivered to a single recipient (i.e., contains no other recipients in either the `to:' or `cc:' fields). For 131 (32.3\%) of these emails, participants indicated that they would be uncomfortable with meeting attendees seeing the resultant email notifications. This number falls to 23.2\% for emails delivered to multiple individuals. This difference is statistically significant ($Z=3.228$, $p=0.001$), suggesting that, when multiple people are in an email thread, the likelihood of the email containing sensitive or private information may be lower.

\textbf{Number of external meeting attendees:} Results from the earlier retrospective survey also suggested that meeting attendance might influence how comfortable people are with sharing their email notifications\footnote{Results from the retrospective survey were not statistically significant, but suggested a possible trend worth further investigation (Figure \ref{fig:meeting_size}).}. Specially, the presence of people outside of one's team or organization \emph{might} increase levels of discomfort. Again, the labeled-data allows for a more ecologically valid and sensitive analysis: Of the 1040 email-meeting pairs, 298 (27.8\%) were attended by people from outside of the organization.\footnote{External to the organization via the \emph{numExternal} feature. Note, the organization chart used in the labeling tool was not sufficiently fine-grained to determine if a fellow employee was a member of a different team.} Participants reported that in 95 cases (32.3\%), they would be uncomfortable sharing the email notification with meeting attendees. This proportion falls to 24.4\% for meetings in which all attendees are fellow employees of the same organization. This difference is statistically significant ($Z = 2.78$, $p=0.005$), suggesting that meeting attendance may indeed influence comfort levels about email notifications.

Together, these findings reinforce and extend our answers to the first two research questions: people are often interrupted by emails when in meetings, in a sizable minority of cases people are uncomfortable sharing the resultant email notifications with the people in the room (RQ1). These levels of discomfort may depend on individual characteristics, as well as on features of the meetings and emails (RQ2). This, in turn, hints at the possibility of using machine learning to predict when email notifications pose an information disclosure risk. We examine this in the next section, and, in doing so, answer our final research question (RQ3).

\section{Predicting Comfort Level}
To further explore the problem space, we developed binary classifiers that, given an email-meeting pair, decide if the delivery of an email notification would result in an uncomfortable situation. In constructing these classifiers, we both: (1) gain a deeper understanding of how combinations of user, email and meeting features may contribute to one's concerns about email notifications, and (2) explore modeling decisions and requirements that can lead to context-dependent predictions accurate enough to be used to manage users' notifications in real-world settings. We address these goals below, with two experiments.

\begin{table}
\centering
\begin{tabular}{|c|l|l|l|l|l|}
 \hline
 & \textbf{Classifier Features} & \textbf{AUC} & \textbf{Pr} & \textbf{Re} & \textbf{F1} \\
 \hline
 \multirow{5}{*}{\rotatebox[origin=c]{90}{Stratified}} &  &  &  &  & \\
 & user & 0.68 & 0.72 & 0.17 & 0.28\\
 & email + meeting & 0.54 & 0.56 & 0.12 & 0.20\\
 & user + email + meeting & 0.76 & 0.64 & 0.45 & 0.53\\
 &  &  &  &  & \\
 \hline
 \multirow{3}{*}{\rotatebox[origin=c]{90}{70:30}} &  &  & &  & \\
 & user + email + meeting & 0.85 & 0.71 & 0.62 & 0.66\\
 &  &  &  &  & \\
 \hline

 \end{tabular}

\caption{Results of the classifiers that predict if a respondent would be comfortable revealing a given email notification in a particular meeting context (i.e., an email-meeting pair). The top section reports results when training and testing on data that is stratified such that no user appears in both collections (i.e., no opportunity for personalization). The bottom section reports results arising when the training and test collections are split such that each user contibutes 7 training examples and 3 test examples (i.e., potential for weak personalization).}
\label{table:classifier_results}
\vspace{-1.5em}
\end{table}

\subsection{Experiment 1: User-Stratified Classification}
In our first experiment, we wish to better understand how \emph{generic} features of people, emails, and meetings impact one's comfort level in receiving email notifications during meetings. This experiment aligns with a cold-start scenario, where a system has no prior labels for a given user. To accomplish this, we stratify our training and evaluation sets such that no user contributes labels to both sets.

To explore the feature space, we train 3 boosted tree ensemble classifiers. We chose to experiment with tree-based classifiers because they are often more interpretable, allowing us to better learn about the design space and understand the importance of features even in a non-linear decision boundary. The first classifier relies only on features about the \emph{user} (e.g., Table \ref{table:features}, top). This model is context-independent, and does not depend on the properties of meetings or emails. A second classifier is trained on \emph{email and meeting} features (Table \ref{table:features}, middle and bottom). A third classifier is trained on all features.

For training and evaluation we use the 1040 labeled examples collected in the second study, above. Training and evaluation is done using 10-fold cross-validation, stratified by users, and optimized to maximize the area under the receiver operating characteristic curve (AUC).

Detailed results are presented in Table \ref{table:classifier_results} (top, stratified). 
There are a few points to note. First, given that the $AUC > 0.5$ in all cases, all three classifiers perform better than chance -- though this is barely the case for the classifier that uses only email + meeting features.
Despite the weak performance of the classifier using email + meeting features, the individual features chosen by the model as informative aligns well with our intuition and with observations from survey and labeled data collection. This feature set contains: the number of external meeting attendees, the number of people on an email's `cc:' line, mentions of money or finance in the mail body, and the presence of negative emotion words in both the email and meeting bodies (Table \ref{table:top_features}).
The user features are more informative than email + meeting features in predicting user comfort, hinting to our observation about three groups of respondents, which can potentially be predicted based on user features such as their role, their rank in the organization and communication patterns.

\begin{table*}
\centering
\adjustbox{width=0.8\textwidth}{
\begin{tabular}{|l|l|l|l|}
\hline
user & email + meeting & all & 70:30 all \\
\hline
U.avgEmlPerMtg (1) & M.numSubjectWords (1) & U.numEmails (1) & U.avgEmlPerMtg (1)\\
U.numMtgWithEml (0.78) & M.numExternal (0.96) & U.avgEmlPerMtg (0.97) & U.numEmails (0.86)\\
U.orgDepth (0.39) & E.numBodyWords (0.95) & E.numBodyWords (0.9) & U.numMtgWithEml (0.78)\\
U.numMeetings (0.36) & E.LIWC[money] (0.83) & E.numSubjectWords (0.82) & M.orgChartAbove (0.73)\\
U.numEmails (0.35) & E.entity[org] (0.79) & E.LIWC[money] (0.79) & E.LIWC[money] (0.7)\\
& E.numCC (0.78) & U.numMtgWithEml (0.77) & M.LIWC[negemo] (0.62)\\
& E.LIWC[negemo] (0.68) & U.orgDepth (0.73) & M.numSubjectWords (0.6)\\
& E.attachment[png] (0.68) & U.numMeetings (0.72) & E.numBodyWords (0.58)\\
& M.LIWC[negemo] (0.67) & E.LIWC[posemo] (0.71) & M.LIWC[posemo] (0.58)\\
& E.numThreads (0.58) & E.numThreads (0.7) & U.numMeetings (0.58)\\
\hline
\end{tabular}}
\caption{Top 10 most informative features for classifiers trained on various feature sets:  user, email + meeting, and all. The leftmost 3 columns show results for the classifier trained on user-stratified data (no personalization). The rightmost column shows results for the full feature set, in the 70:30 split condition (weak personalization). The number in the parentheses indicates the  information gain, normalized such that the most informative feature scores a 1.0. The prefixes 'U', 'E' and 'M', denote user, email and meeting features, respectively.}
\label{table:top_features}
\end{table*}

Although email + meeting features are weakly informative alone, their combination with user features lead to larger predictive performance than each feature set achieves alone. We present the relative importance of features in the combined model in the third column of Table \ref{table:top_features}. Here, we found that the most informative features are related to how many emails a user receives, and the average number of emails they receive during a meeting. This aligns with our general intuition that the more emails a user receives when they are in a meeting, the greater risks they would be exposed to potential information disclosure events. Likewise, and likely for the same reasons, we find that the number of meetings interrupted by email, and the total number of meetings attended per week are also predictive of information disclosure risk. Another set of informative features is related to email length. Specifically, the length of the email subject and body, and the number of emails in the email thread were found to be important. Finally, we found that mentions of finance and instances of positive emotion words were important features. This aligns with previous research suggesting that people are uncomfortable sharing information in certain categories such as money ~\cite{olson_chi05}.

\subsection{Experiment 2: Personalization}
Given the importance of user-level features above,
we now explore the gains that can be achieved by allowing the classifier to learn from a user's prior labels. Specifically, we retrain our best performing model (user + email + meeting features), but stratify the data such that, for each user, 7 labeled data points are included in the training set, and 3 labeled data points are included in the test split. This arrangement allows for a weak form of personalization. We again use 10-fold cross-validation for the evaluation.

Results are presented in the bottom row of Table \ref{table:classifier_results} (70:30). This model outperforms the previous best-performing model by 12.0\% (as measured by AUC). This result strongly argues for the need to personalize models of notification information disclosure risk. With our best performing model achieving a precision of 0.71, and a recall of 0.62, the classifiers are likely to be too weak useful in most high-risk applications. Nevertheless, our explorations reveal that generic features provide some information about the information disclosure risks of an email-meeting pair (i.e., this model does substantially better than guessing). 

Similar to Experiment 1, we also looked at the 10 most informative features with this classifier (the right most column in Table \ref{table:top_features}). In addition to the user features that might be indicative of potential exposure (the number of emails a user receives in a week, the average emails per meeting, the number of meetings), and email features that are related to email contents and length, we found features related to meeting context. Notably, the third most informative feature was one that counted how many meeting attendees were above the recipient in the organizational chart. 
Again, we also found that people were more uncomfortable sharing emails when the meeting description contains strong negative or positive emotion words.

To this end, we have answered our research questions. In the next section, we offer design implications, then conclude with a brief discussion of limitations and future work.

\section{Design Implications}
In this section, we discuss some of the implications for designing contextually-relevant notification systems and policies.

\textbf{Personalization matters.} Both our exploratory retrospective survey and the contextual labeling study revealed three general and distinct groups of respondents. If a user can be quickly characterized as \emph{Always Comfortable} or as \emph{Always Uncomfortable}, then the notifications policies are rather straightforward: unconditionally allow all notifications for the unconcerned, and turn off notifications for those \emph{Always Uncomfortable} unless the system is certain the user is alone. However, there exists a much richer strategy space for the \emph{Mixed Comfortable} group, where various notification actions can be designed by hiding certain fields of messages, or delivering notifications only to certain devices.

\textbf{Even simple context is helpful.} As discussed above, for some people an effective notification policy might need only know if a person is alone, and examining a user's calendar may serve as an acceptable approximation. Slightly more sophisticated policies might instead consider whether an email was delivered to multiple people, or whether a meeting will be attended by people from outside the organization. These possibilities make clear that interesting notification policies can be developed from simple, easy-to-compute, signals.

\textbf{Better sensing is likely to help.} Both the retrospective survey and the contextual-labeling study included questions that could not be automatically answered based on calendar appointments and emails alone. For example, we asked participants if they actually attended meetings, and whether meetings were conducted via teleconference. With richer sensing capabilities, it is possible that these features too could be reasoned about automatically.

\textbf{Choose your notification fields wisely. } Finally, our studies revealed that certain notification fields are generally less sensitive than others. However, these preferences weren't universal, and varied even within a single user's responses. For example, an email's sender field may pose little risk in some cases (e.g., an email from a known collaborator), but pose significant risk in others (e.g., an email from a particular medical specialist). Systems should leverage these general trends and preferences, but also be expressive enough to allow exceptions. In addition, to better preserve user privacy while deciding which notification fields to show, we not only need to take into consideration the sensitivity of different notification fields in various context and contents, but also need to use rather more conservative notification strategies in disclosing information when there is uncertainty in predicting comfort level.

\section{Discussion and Limitations}
In this paper, we characterized information disclosure risks that arise when people receive notifications in the presence of others. Specifically, we focused on emails that arrive during meetings. Our analysis employed multiple lines of evidence, including one large retrospective survey, and data collected in a second study via a purpose-built labeling tool. We report similar trends in both datasets, and are encouraged by the consistency of the findings.

Nevertheless, we caution readers against overgeneralizing our findings. Both the survey and labeling tool were deployed within a single large U.S.-based information technology company. Though respondents occupied a wide variety of job roles, it remains to be demonstrated that our findings generalize to other companies, company cultures, and industries. For example, it seems likely that people working in financial, legal, and medical industries may be more sensitive to information disclosure risks. It is also possible that preferences and concerns may vary by country and culture. Thankfully the \emph{features} we described in this paper are very general, and are likely to be applicable in other companies and industry settings. As such, we expect replication efforts to be rather straightforward, and we both encourage and anticipate efforts to replicate our findings in other industries and cultures.

Additionally, we would like to extend our analysis to personal emails and contexts (e.g., social gatherings, appointments, etc). The retrospective survey revealed that people perceive a higher risk when personal emails arrive during work meetings. One wonders: Is the same true for social gatherings? And, how might work notifications be perceived in non-work contexts? We believe this line of investigation will lead to fruitful future research.

Likewise, our studies focused exclusively on email notifications. This choice was deliberate and practical; as noted in the introduction, email is often a carrier for other types of notifications (e.g., social networks), and by focusing on emails we were able to build and deploy a labeling tool with minimal effort. In the future, we hope to explore the privacy risks of other notification types, including: instant messages, calendar appointments, to-dos, personal reminders, and other information proactively displayed by virtual assistants. Studying these notifications will require deeper technical integration with notification-capable devices and platforms.

Our studies were also exclusively performed from the perspective of the notification recipient. We recognize that the senders of emails are also subject to disclosure risks. When such risks are present, message originators can take some limited preventive actions to mitigate risks (e.g., by indicating that an email is private or sensitive in email subject, or by adding blank lines to the beginning of a message). In the future, we hope to investigate these aspects of the problem space and potentially explore practices that can be shared across organizations to mitigate such risks.



On a technical note, we also reflect on our limited success in constructing a classifier for meeting-email pairs. As noted earlier, our models were trained and tested on a total of 1040 labeled pairs. While this sample is large in the context of a survey, and for reporting descriptive statistics, it is rather small for state-of-the-art classifiers. This data scarcity is confounded by the fact that labels within the dataset are likely correlated; the 1040 labeled pairs represent only 169 individuals. In the future, we hope to collect more data and improve prediction accuracy to implement real-time notification-capable systems.

Finally, we recognize that there are costs to delaying notifications, or hiding notification fields \cite{horvitz_cacm03, bayesphone}. Our studies did not measure these costs, and cannot directly experiment with formal notification policies. Nevertheless, we feel our analysis has showcased the need and opportunity to develop such policies, and has provided strong hints about which features and properties are likely to be informative.

\section{Conclusion}
In this paper, we report results from an exploratory retrospective survey and a larger contextual labeling study. Our research  necessitated that we ask users to discuss sensitive scenarios. Specifically, we asked user to describe sensitive emails, and to discuss why they would feel uncomfortable receiving notifications for those messages when in the presence of others.  To this end, we designed our studies to carefully respect participant privacy, and we believe these considerations were instrumental in allowing us to recruit a combined total of 300 individuals. From these individuals, we learned:

\begin{itemize}
\item (RQ1) Email notifications indeed pose an information disclosure risk.
\item (RQ2) The real or perceived severity of these risks depend both on user characteristics (e.g. the nature of occupation) and attributes of the meeting or email (e.g., the number of recipients or attendees).
\item (RQ3) Machine-learned models can learn attributes, patterns and signals associated with risky email-meeting pairs. Here, user-level features are more informative than generic meeting or email-level features. The best performing models incorporate all features, together with user history.
\end{itemize}

Taken together, our findings present a rich picture of notifications, as viewed through the lens of privacy. We hope that our findings will inform the design of future notification-capable systems.

\bibliographystyle{SIGCHI-Reference-Format}
\bibliography{kamloops,privacy2}
\end{document}